# Nuclear emulsions in the OPERA experiment

Donato Di Ferdinando[a] for the OPERA Collaboration[1]

[a]*Istituto Nazionale di Fisica Nucleare, Viale Berti Pichat 6/2, Bologna, Italy*

**Abstract**

The use of emulsions as particles detector has a long and successful life. The recent development of fast automatized analysis systems has allowed the use of huge amount of emulsions films with no precedence in the history of particle physics. The OPERA experiment, running in the underground Laboratori Nazionali del Gran Sasso (LNGS), for the confirmation of the neutrino oscillation in the $\nu_\mu \longrightarrow \nu_\tau$ channel, is "the major" of these experiments. The experimental technique, the strategy and the first results of the experiment are presented.

## 1. Introduction

The neutrino oscillation hypothesis explains the long standing solar and atmospheric neutrino puzzles. In this theory neutrinos have non vanishing masses and their flavor eigenstates involved in weak processes, decays and reactions are a superposition of their mass eigenstates [Pontecorvo 1957, 1958, Maki et al.1962].

In the last decades several key experiments conducted with solar, atmospheric, reactor and accelerator neutrino allowed to build up our present understanding of neutrino mixing. In particular, the first studies on neutrino oscillations, based on atmospheric neutrino, were done by Kamiokande [Fukuda 1994], SuperKamiokande [Hirata 1988, Fukuda 1998], MACRO [Ahlen 1995, Ambrosio 1998, Ambrosio 2003, Ambrosio 2004] and SOUDAN II [Allison 1999, Allison 2005]. Accelerator based neutrino experiments, K2K [Ahn 2006] in Japan and MINOS [Gallagher 2008] in USA, confirmed the oscillation hypothesis by measuring the $\nu_\mu$ flux deficit at large distances and its energy dependence in the atmospheric mass range of 2-3 $10^{-3}$ $eV^2$. The CHOOZ [Apollonio 2003] and Palo Verde [Piepke 2002] reactor experiments excluded the $\nu_\mu \longrightarrow \nu_e$ as the dominant one. However, the appearance of a different neutrino flavor is still an important issue. The Oscillation Project with Emulsion tRacking Apparatus (OPERA) experiment [Shibuya 1997, Guler 2000, Declais 2002] was conceived to conclusively confirm the $\nu_\mu \longrightarrow \nu_\tau$ oscillations in the parameter region indicated by the atmospheric neutrino analyses through direct observation of $\nu_\tau$ appearance in a pure $\nu_\mu$ beam.

## 2. The OPERA experiment

A muon neutrino beam, produced at CERN, is fired to the OPERA detector located in the underground Gran Sasso laboratory, 730 km away. The challenge of the experiment is to measure the appearance of $\nu_\tau$ from $\nu_\mu$ oscillations. This requires the detection of the short-lived $\tau$ lepton ($c\tau = 87.11$ μm) produced in the charged-current (CC) interaction of a $\nu_\tau$. The $\tau$ is identified by the detection of its characteristic decay topologies, in one prong (electron, muon or hadron) or in three-prongs. To cope with the requirements of large mass detector (due to low neutrino cross section) and high tracking resolution (due to the short life of $\tau$ lepton) the Emulsion Cloud Chamber (ECC) technique was used in OPERA.

## 3. The CNGS beam

The CERN to Gran Sasso (CNGS) neutrino beam [CNGS project] was designed and optimized to maximize the number of CC interactions of $\nu_\tau$ produced by oscillation at the Gran Sasso underground laboratory.

A 400 GeV proton beam is extracted from the CERN SPS in 10.5 μs short pulses with a design intensity of 2.4 x $10^{13}$ protons on target (pot) per pulse. The interactions of the proton beam with a series of thin graphite rod helium-cooled produces secondary pions and kaons that, by a system of two magnetic lenses (the horn and the reflector) are focused into a quasi-parallel beam. In the 1000 m long decay pipe the pions and kaons decay mainly in muons and muon-neutrinos. Detectors

---
[1] For a list of names and Institutions of the OPERA Collaboration See ref. A. Anokhina et al. JINST3, P07002, 2008.

at the end of the dump allow monitoring the flux of μ and $\nu_\mu$, that are pointed toward the Gran Sasso. The muons are absorbed in the rock while neutrinos continue their travel toward the detector (Fig.1)

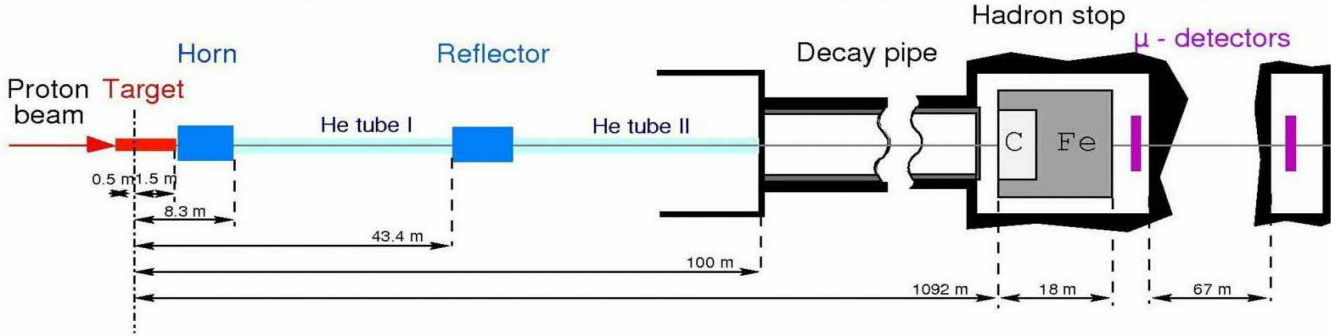

Figure 1. Schema of CNGS beam

The average neutrino energy is 17 GeV. The anti-neutrino muonic contamination is around 4%, the $\nu_e$ contamination is lower than 1% and the number of prompt $\nu_\tau$ is negligible. In one year the nominal integrated beam intensity is $4.5 \times 10^{19}$ pot.

### 4. OPERA detector

The OPERA detector is a large ($10 \times 10 \times 20$ m$^3$) hybrid detector made of passive and active components.

The passive component is contained in the Emulsion Cloud Chamber (ECC). The basic ECC unit of OPERA detector is the *brick* that consists of a sequence of 57 emulsion films interspaced with 56 1 mm thick lead plates packaged in a light tight and high mechanical precision way. One OPERA brick weights 8.2 kg, and has a thickness along the beam direction of 7.5 cm, equivalent to 10 radiation lengths.

The bricks are arranged in planar vertical structures called walls, with transverse dimensions of about $6.7 \times 6.7$ m$^2$. Each brick wall is coupled with a wall of Target Tracker (TT). On each downstream face of a brick is attached a box containing two emulsion films, called Changeable Sheets (CS); they are the interface between the TT detectors and the tracks recorded in the bricks [Anokhina 2008a]. A sequence of 31 TT interleaved with the brick walls compose the target section. The latter together with a downstream muon spectrometer form a supermodule. OPERA consists of two identical supermodules and a veto system (planes of glass Resistive Plate Chambers, (RPC) in front of the first supermodule) to tag the interactions occurring in the surrounding rock of the experimental hall (Fig.2). The total number of bricks hosted in the two target sections is 154000 for a global mass of about 1.3 kton.

#### 4.1 Electronic Sub detectors

Each TT wall consists of two planes of plastic scintillator strips (6.8 m x 2.6 cm x 1 cm) one providing the vertical and

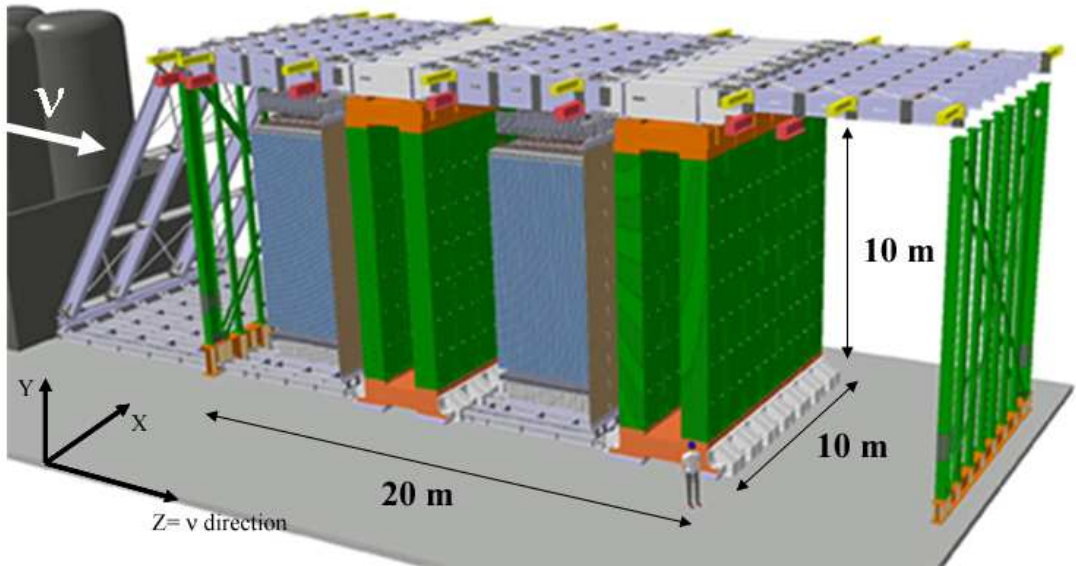

Figure 2. Schematic side view of the OPERA detector.

The active component of the detector is made by electronic detectors.

the other the horizontal coordinates. Each plane of TT is made of 4 horizontal and 4 vertical modules. Each module consists of 64 scintillator strips with a wavelength scintillating fiber (1 mm diameter) glued on each strip. The signal from the fibers is readout, at both ends, by a 64-pixel photo multiplier. The TT provides the trigger and identifies the brick in which neutrino interaction has taken place. The detection efficiency is 99%. [Adam 2007].

Each OPERA spectrometer consists of a dipolar magnet (8.75 ×8 m$^2$) made of two iron arms producing a field of 1.52 T. Each arm is composed of 12 iron slabs interleaved with bakelite RPC's providing a coarse tracking inside the magnet, range measurement of the stopping particle and calorimetric information [Ambrosio 2004, Bergnoli 2007]. The spectrometer is equipped with a High Precision Tracker (HPT), placed in front, in the middle and behind each dipole magnet. HPT are composed of six fourfold layers of vertical planes of 8 m long drift-tubes with an outer diameter of 38 mm and a sense wire of 45 μm. The HPT, that has a spatial resolution of 300 μm, allows precision measurements of the muon momentum and determines the sign of their charge with high accuracy, $\Delta p/p < 25\%$ and charge misidentification below 1% for p<25 GeV/c [Zimmermann 2005a, 2005b]. In order to solve ambiguities in the track reconstruction, two planes of RPC with crossed readout strips (45°) are placed upstream of each dipole magnet to roughly complement the information coming from the drift tubes.

The data from the muon spectrometer combined with those of TT provide a muon identification probability of at least 95%.

## 5. OPERA emulsion film and the ECC concept

The 154000 OPERA brick comprise ~9 millions nuclear emulsion films, the largest amount of emulsion ever used in high-energy physics experiments.

In the past experiments, the emulsions were poured by hand following standard procedures developed in many years of experience. The same procedure applied to OPERA would have been prohibitively time consuming. An R&D project carried out by the Nagoya University and the Fuji Film co. has allowed solving this problem by using dedicated commercial photographic film production lines to automatize the coating of emulsions. This solution presents the advantage to precisely control the thickness of the emulsion, impossible to do with manual film production.

The OPERA emulsion film is 10.2 x 12.5 cm$^2$ and is composed of a 205 μm of plastic base with a 44 μm of sensitive layer on both face of the base (Fig.3); It has an AgBr content higher with respect to the commercial photographic emulsions. The diameter distribution of AgBr crystals in the emulsion layer is also more uniform with a peak at 0.20 μm. The sensitivity is 30 grains/100 μm for a Minimum Ionizing Particle (MIP). Because of their continuous sensitivity, the emulsion films, after the production, are refreshed[2] in a dedicated underground facility at the Tono mine (Japan) in order to remove latent tracks due to cosmic rays and ambient radioactivity.

---

[2] The refreshing operation consists in accelerating the fading process. For the OPERA emulsions this is obtained keeping the films at HR=95% and Temperature of ~20-29 °C for 3 days.

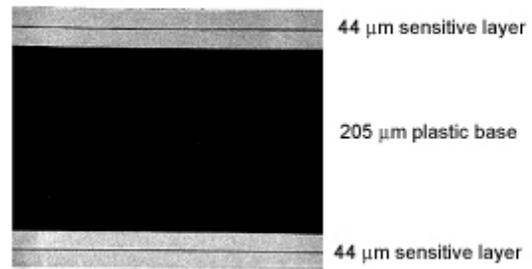

Figure 3. Electron microscope micrograph of a section of an OPERA Emulsion film.

The ECC technique allows building a modular detector that has high mass (lead) and high spatial and angular resolutions (emulsion).

There are two ways in which a τ lepton may decay in OPERA brick. When the τ decays in the same lead plate where the neutrino interaction occurred, the τ candidates are selected on the basis of the Impact Parameters (IP) of the τ daughter tracks with respect to the interaction vertex. The vertex and thus the IP can only be determined reliably for multi-prong deeply inelastic scattering (DIS). When τ decays in the first or second downstream lead plate it is detected by the angle between the charged decay daughter and the parent direction both for DIS and quasi elastic (QE) neutrino interactions. In addition the characteristics of the OPERA bricks allow measuring the hadrons momenta from their multiple Coulomb scattering, separating low energy pions and muons from their different energy loss, identifying electron and measuring the energy of electrons and photons [De Serio 2003, Arrabito2007b].

## 6. OPERA strategy.

The neutrinos coming from CERN cross the target section of the detector. The TT triggers the interaction and identifies the brick most probably containing the event. The brick is removed from the target wall by two robots forming the Brick Manipulator System (BMS). Then a first reference frame, made by 4 X-rays spot, is printed between the CS doublet and the first downstream plate of the brick. The CS doublet is detached from the brick and developed in the underground facility. In the scanning station a connection between the TT and the CS doublet tracks is searched for. If there is no connection, a new CS doublet is attached to the brick, and the brick is put back in the detector. If the connection is found, a further lateral X-ray reference frame is marked on all the emulsion films of the brick to speed up the scanning procedure.

Then the brick is brought to surface and exposed for 12 hours to cosmic rays. The exposition is done in a specially designed pit, shielded from soft lateral cosmic rays, in order to mainly select muons passing through the entire brick, with a density of 100 tracks/cm$^2$ necessary for film-to-film alignment. The brick is then disassembled, developed in an automatized facility and sent to the scanning laboratories in Europe and Japan.

## 7. Emulsion Scanning

When a charged particle hits an emulsion layer, it ionizes the medium along its path, leaving, after development, a sequence of aligned silver grains, with linear size of about 0.6 μm. With a high magnification optical microscope, silver grains appear as black spots on bright field. The large emulsion surface to be scanned requires fast automated microscopes running at speeds of at least 20 cm$^2$/h keeping the good resolution provided by nuclear emulsions (< 1μm). After R&D studies conducted using two different approaches, the "European Scanning System" [Armenise 2005, Arrabito 2006, Arrabito 2007] and the "S-UTS" [Nakano 2008] in Japan, this requirement was successfully achieved. Although the two scanning systems are quite different in the design and realization, the measuring philosophy is the same: to reconstruct a particle trajectory, the computer driven microscope adjusts the focal plane of the objective lens through the whole thickness of the emulsion, obtaining an optical tomography at different depths of each field of view, typically one image every 3 μm. Each image is filtered and digitized in real time to recognize track grains as clusters of dark pixels. A tracking algorithm reconstructs 3D sequences of aligned grains and extracts a set of relevant parameters for each sequence. The position accuracy obtained for the reconstructed tracks is about 0.3 μm with a resulting angular resolution of about 2 mrad.

The vertex finding strategy consists in following back, film by film, the tracks found on the CS doublet, till the track stops inside a lead plate in the brick. This is the so called *scan-back* method sketched in Fig.4. To confirm the stopping track, an area scan of several mm$^2$ around the last found segment of the track is performed for 11 plates (typically the plate containing the stopping segment, 5 plates upstream and 5 plates downstream). The acquired areas are then aligned (using the passing through cosmic rays tracks) and all the tracks in the volume are reconstructed. Vertex algorithms consider only those tracks pointing to the stopping segment, a vertex interaction is reconstructed and a topology compatible with the decay of a τ lepton is searched for.

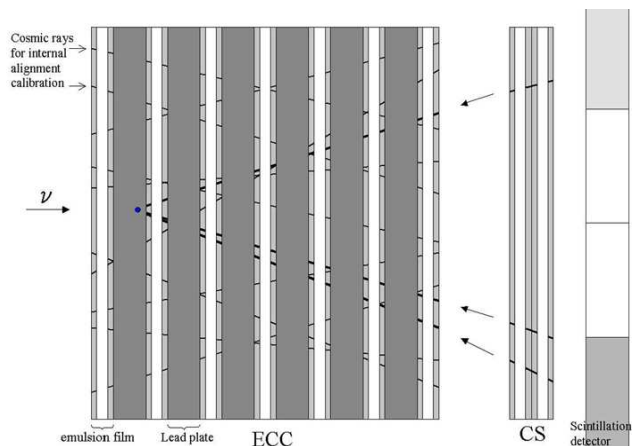

Figure 4. Scheme of the scan-back method. The tracks, founded on the CS doublet and matched with the TT, are followed back till the interaction point.
Then a volume scan around the stopping track is analyzed in order to reconstruct the vertex and confirm the interaction.

For five years of data taking with a nominal beam intensity of 4.5 x 10$^{19}$ pot/year, the expected number of τ events is 10.4 for $\Delta m^2$ = 2.5 x 10$^{-3}$ eV$^2$. The number of signal events essentially scales like $(\Delta m^2)^2$. The background is expected to be less than one event.

**8. OPERA performances**

The first time in which the OPERA detector was exposed to neutrinos coming from CERN, took place in August 2006. A low intensity beam lasted 13 days and provided a total luminosity of 7.6 x 10$^{17}$ pot, equivalent to 5 days of running at nominal intensity. At that time, there were no bricks in the target section and only electronic detectors were installed together with a plane of changeable sheets; using the time correlation between the beam and the event recorded in the detector 319 events were selected out of 300 expected. In this first technical run the synchronization between the GPS clocks between CERN and Gran Sasso were fine tuned; the full reconstruction of the electronic detectors data were tested and the average zenith angle of penetrating muon tracks measured by the detector was (3.4 ± 0.3) ° in very good agreement with the value of 3.3° expected from the direction of the CNGS beam. Further the connection between the electronic detectors (few cm of resolution) and CS emulsion films (few μm) was established for the first times, Fig.5 [Acquafredda 2006].

The first physics run was held in October 2007 with 40% of target section filled with bricks. The run was very short in duration, due to failures of the ventilation control units of the proton target, caused by high radiation level in the area were the units were installed. Consequently an integrated luminosity of 0.82 x 10$^{18}$ pot was accumulated, corresponding to 3.6 day at nominal intensity. Notwithstanding the beam problems, the detector was fully operational and an interaction vertex was successfully reconstructed in the first brick extracted (Fig.6).

In total 38 events were triggered in the OPERA detector to be compared with 32 ± 6 expected. Despite of its short duration the run allowed a successful testing of the electronic detectors, data acquisition and brick finding algorithms; it proved the ability of the matching between the target tracker and the bricks and validated the full scanning strategy.

Fig.6 shows the picture of the first event recorded; it is a Charge Current (CC) interaction with 5 prongs and a shower pointing to the primary vertex (γ conversion after a π$^0$ decay).

In Fig.7 an example of a Neutral Current candidate event is displayed.

The 2008 run has started in June and till the end of August, half way of the run, OPERA has accumulated 4.5x10$^{18}$ pot, 60% less than expected. This was due to accidents that have affected the beam during the first weeks. However the OPERA detector has registered 399 events in the brick walls out of (434±21) events expected[3].

**9. Conclusion**

The OPERA detector received the CNGS ν beam for a brief period in 2006 and obtained the first interaction events in a second short run in 2007. The runs allowed to successfully testing the experiment strategy, from brick finding to event location in the emulsions. The detector is now essentially filled with bricks. The 2008 run, started in June, had an initial period in which the beam suffered a series of problems but it has since then considerably improved both in stability and intensity.

While I m writing this paper neutrinos are passing under my feet and some interact in the detector. Hope that the first $\nu_\tau$ will interact in the following days.

---

[3] At the moment of writing this document, on 12 October 2008, the beam has significantly improved both in stability and in intensity, a total of 1.23 x 10$^{19}$ pot have been accumulated, almost 60 % of what was anticipated.

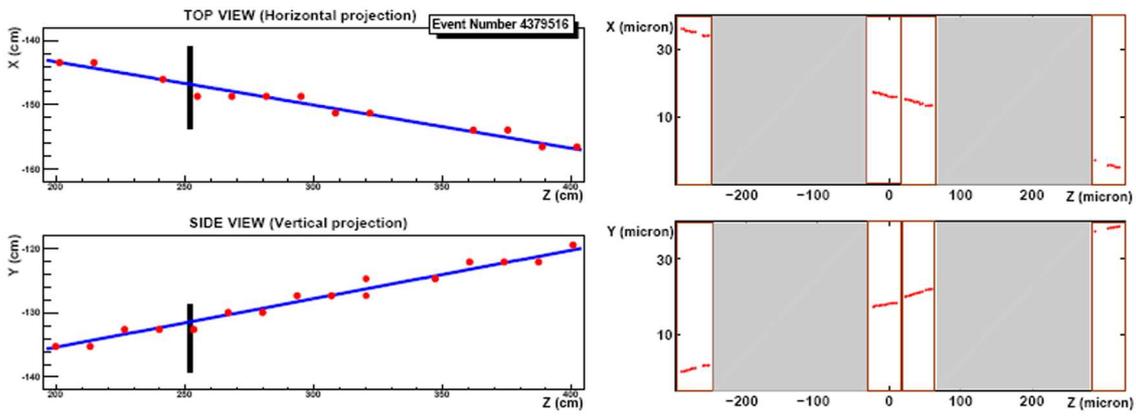

Figure 5. Left: display of one event with the muon passing through the CS detector plane. Only hits of the electronic detectors close to the CS plane are shown; the vertical segments indicates the position of the CS doublet intercepted by the track. Right: display of the corresponding tracks reconstructed in the CS doublet.

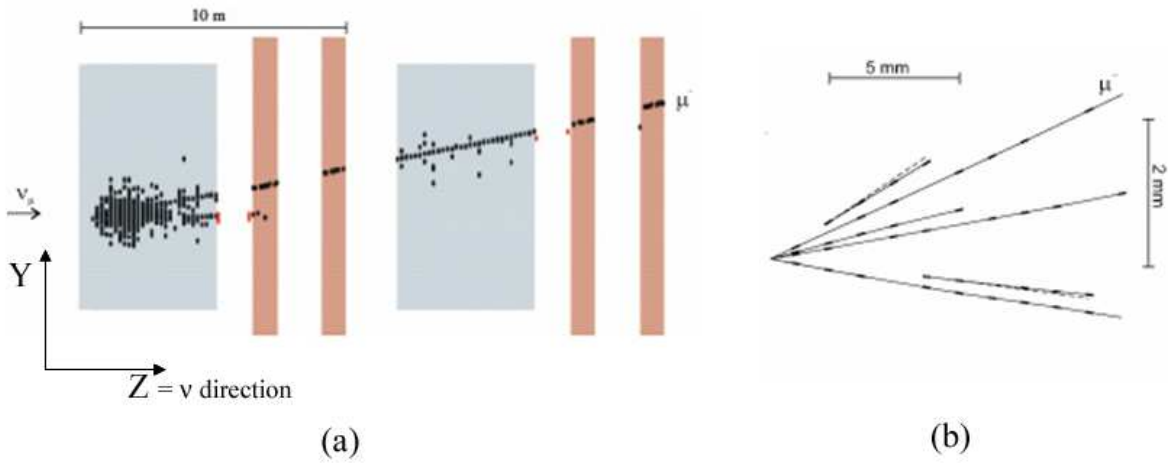

Figure 6. First CC event recorded by OPERA detectors. Left: side view of the event as seen by the electronic detectors. Right: Side view of the event reconstructed in the emulsions.

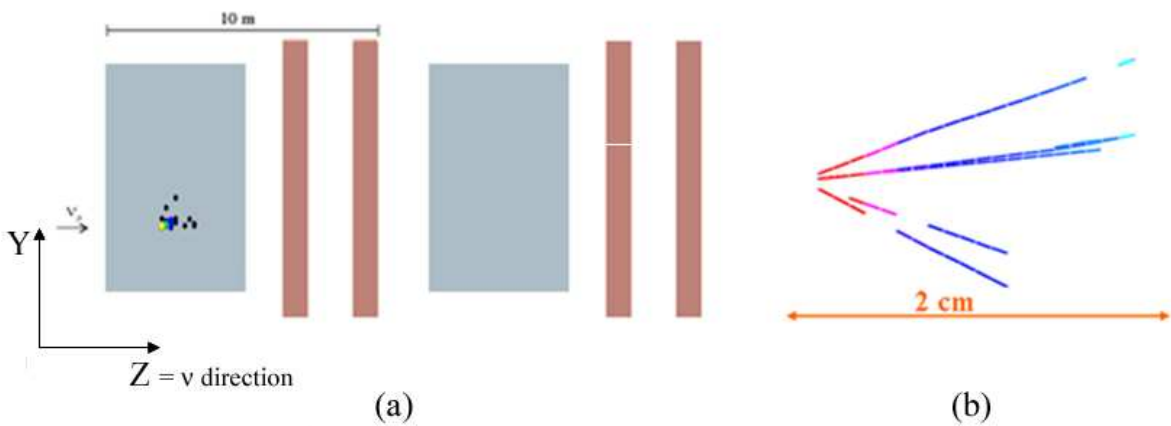

Figure 7. Neutral Current candidate as reconstructed in electronic detector (a) and emulsions (b).